\documentclass[twocolumn]{emulateapj}  
\usepackage{epstopdf}
\usepackage{ulem}
\usepackage{amssymb}
\usepackage{natbib}
\usepackage{times}
\usepackage{graphicx,bm,amssymb}
\usepackage{pstricks}

\usepackage{psfrag}
\usepackage[colorlinks=true,linkcolor=blue,citecolor=blue]{hyperref}
\newcommand{\be}{\begin{equation}}
\newcommand{\ee}{\end{equation}}
\newcommand{\bse}{\begin{subequations}}
\newcommand{\ese}{\end{subequations}}
\newcommand{\bary}{\begin{eqnarray}}
\newcommand{\eary}{\end{eqnarray}}

\shorttitle{}
\shortauthors{Fraija N.}
\begin{document}
\title{The origin of the optical flashes: The case study of GRB 080319B and GRB 130427A}
\author{N. Fraija$^{1}$ and  P. Veres$^2$}
\affil{$^1$ Instituto de Astronom\' ia, Universidad Nacional Aut\'onoma de M\'exico, Circuito Exterior, C.U., A. Postal 70-264, 04510 Cd. de M\'exico,  M\'exico.\\
$^2$ Center for Space Plasma and Aeronomic Research (CSPAR), University of Alabama in Huntsville, Huntsville, AL 35899, USA
}

\email{nifraija@astro.unam.mx and pv0004@uah.edu}

\date{\today}
\begin{abstract}
Correlations between optical flashes and gamma-ray emissions in gamma-ray bursts have been searched in order to clarify the question whether these emissions occur at internal and/or external shocks.  Among the most powerful gamma-ray bursts ever recorded are GRB 080319B and GRB 130427A which at early phase presented  bright optical flashes possible correlated  with $\gamma$-ray components. Additionally,  both bursts were fortuitously located within the field of view of the TeV $\gamma$-ray Milagro and HAWC observatories, and although no statistically significant excess of counts were collected, upper limits were placed on the GeV - TeV emission.   Considering the synchrotron self-Compton emission from internal shocks and requiring the GeV-TeV upper limits we found that the  optical flashes and the $\gamma$-ray components are produced by different electron populations.   Analyzing the optical flashes together the multiwavelength afterglow observation, we found that these flashes can be interpreted in the framework of the synchrotron reverse-shock model when outflows have arbitrary magnetizations. 
\end{abstract}


\maketitle

\section{Introduction}
Gamma-ray bursts (GRBs) are classified as one of the most energetic events in the universe.  Our understanding of GRBs has  improved significantly in the last 15 years. Observations have firmly established that GRB prompt phases and their afterglows arise from highly relativistic and collimated outflows \citep{2002ApJ...571..779P, 2004ApJ...609L...1T}. Based on photometric and spectroscopic observations, long GRBs (lGRBs) have usually been associated  to the core collapse of massive stars \citep{2006ARA&A..44..507W, 2012grbu.book..169H, 2003Natur.423..847H} and short GRBs (sGRBs)
to  the merger of compact objects \citep{1989Natur.340..126E, 1992ApJ...395L..83N, 2004ApJ...608L...5L,2005ApJ...632..421L}.\\
Among the fundamental questions that are not answered yet is the physical origin of the prompt emission in GRBs.   Although it is still uncertain, a typical GRB prompt spectrum is nonthermal and generally well modelled by a so-called Band function \citep{1993ApJ...413..281B} which depends on three parameters: a low-energy power-law index $\alpha$, a high-energy power-law index $\beta$, and a spectral break energy $E^{\rm obs}_p$, which defines the smooth transition between the two power laws.  The observations of GRB prompt emission with low-energy spectral slopes that are inconsistent with synchrotron \citep{1997ApJ...488..330C, 2000MNRAS.316L..45G, 1998ApJ...506L..23P, 2002ApJ...581.1248P} supply further information to consider inverse Compton scattering.    \cite{2016ApJ...816...72Z} suggested that for the origin of the observed GRB prompt emission, at least some, if not all, Band-like GRB spectra with typical parameter values can be interpreted as synchrotron radiation from accelerated electrons in shock waves. Such internal shocks (IS) occurred if the ejection process by the central source is highly variable \citep{1994ApJ...430L..93R, 2004ApJ...613..448P, 2000ApJ...544L..17P, 2007MNRAS.380...78G, 2003ApJ...585..885G, 2007ApJ...671..645A, 2008ApJ...689.1150A}. The observation of high-energy spectral components in  GRB prompt phase can provide strong constraints on present models.  Some authors have claimed that the high-energy emission with photons $>100\, {\rm MeV}$ during the prompt phase (before T$_{90}$) has an internal origin similar to its lower counterpart energies \citep{2011MNRAS.415...77M, 2011ApJ...730..141Z, 2011ApJ...733...22H,  2011ApJ...730....1L}.\\
Optical flashes have been widely discussed in the literature \cite[see][]{2015PhR...561....1K}.  Using standard assumptions such as  the forward- and reverse-shocked shells carry comparable energy,  optical flashes are described by synchrotron emission from reverse shock (RS) which is shown as a single peak \citep{2000ApJ...536..195C, 2005ApJ...628..315Z, 2003ApJ...595..950Z, 2003ApJ...597..455K, 2000ApJ...545..807K},  although if the central engine emits slowly moving material the RS could survive up to weeks \citep{2007MNRAS.381..732G, 2007ApJ...665L..93U}.   \cite{1997ApJ...476..232M} proposed that optical flashes might result from IS even if these flashes from IS are nearly two orders of magnitude weaker than those due to the RS. The authors showed that, with beaming factors of $\sim 10^{-2}$, it is possible to have flashes as bright as 9$^{\rm th}$ magnitude for $z\sim 1$ \citep{1999MNRAS.306L..39M, 2000ApJ...532..286K}.   Considering the FS propagating into the pre-accelerated and pair-loaded environment, some authors have proposed that FS at its early stage could explain the early optical emission \citep{2002ApJ...565..808B, 2001ApJ...554..660M}.   Early observations of GRB afterglows would offer to clarify the question whether the early emission takes place at IS or ES. \\ 
Prompt observations in the optical frequencies remain difficult due to a lack of good temporal coverage.   However,  a strong evidence in favor of a bright additional component at low energies is given by GRB 080319B and GRB 130427A. These bursts are among the brightest and most energetic GRBs which were observed by several satellites and ground-based instruments.    \cite{2008Natur.455..183R}  found that  for GRB 080319B both optical and gamma-ray bands were mildly correlated, leading to both emissions were originated in the same physical region, and  \cite{2014Sci...343...38V} reported that the optical and very-high-energy (VHE) gamma-ray emissions showed a close correlation during the first 7000~s.\\
In this paper, we analyze the origin of optical flashes present in the light curve (LC) of GRB 080319B and GRB 130427A. We consider the GeV - MeV$\gamma$-ray, X-ray and optical data together with the GeV - TeV upper limits derived by  Milagro and HAWC experiments to constrain the  synchrotron self-Compton (SSC) models from IS and early-afterglow ES.  The paper is arranged as follows: in Section 2 we give a brief description of GRB 080319B and GRB 130427A observations; in Section 3 we present a model based on IS and early-afterglow ES  to fit data;  in section 4  we discuss our results, and brief conclusions are given in section 5.\\
\section{Properties of GRB 080319B and GRB 130427A}
In the following subsections we present a brief description of the observations performed around GRB 080319B and GRB 130427A.
\subsection{GRB 080319B}
On 2008 March 19, one of the brightest and most energetic bursts, GRB 080319B, was observed by several satellites and ground-based instruments.   The \textit{Swift}-Burst Alert Telescope (BAT; 15-350 keV) triggered on GRB 080319B at $T_0$ = 06:12:49 UT\citep{2008GCN..7427....1R}.  The burst direction was within the field of view of the BAT for 1080 s, placing strong limits on any precursor emission. This burst was simultaneously  detected with the Konus gamma-ray detector (20 keV - 15 MeV) onboard the \textit{Wind} satellite \citep{2008GCN..7482....1G}. Both Swift-BAT and Konus-Wind (KW) LCs showed a complex and a strongly energy-dependent structure, lasting approximately 57 s \citep{2008Natur.455..183R}.
\noindent The time-averaged KW gamma-ray spectrum was well fit using a Band function \citep{1993ApJ...413..281B}, with $\alpha = -0.855_{-0.013}^{-0.014}$, $\beta = -3.59_{-0.62}^{+0.32}$, and $E^{\rm obs}_p = 675 \pm 22$ keV ($\chi^2 /dof = 110.4/80$).  This burst had a peak flux of $(2.26 \pm 0.21)\times10^{-5}\, {\rm erg\, cm^{-2}\,s^{-1}}$, a gamma-ray fluence of $(6.13 \pm 0.13)\times10^{-4}  \,{\rm erg\, cm^{-2}}$ and an isotropic equivalent gamma-ray energy released of $1.3\times10^{54}\, {\rm erg}$ in the energy range of 20 keV -€" 7 MeV \citep{2008Natur.455..183R}.\\
\noindent The wide-field robotic optical telescope Pi of the Sky€ \citep{2007Ap&SS.309..531C,2008GCN..7439....1C}, and the wide-field robotic instrument Telescopio Ottimizzato per la Ricerca dei Transienti Ottici RApidi (TORTORA, which is attached to the 60 cm robotic optical/near-infrared Rapid Eye Mount \citep[REM;][]{2001AN....322..275Z} telescope located at La Silla, Chile), coincidentally had this burst within their fields of view at the time of the explosion \citep{2008GCNR..121....1P}.  Pi of the Sky observed the bright optical transient from $T_0+2.75$ s to it faded below threshold to $\sim 12^{th}$ magnitude after 5 minutes \citep{2009ApJ...691..723B}.  TORTORA measured the brightest portion of the optical flash with high time resolution enabling to do detailed comparisons between the prompt optical and gamma-ray emission \citep{2008Natur.455..183R}.
\noindent The \textit{Swift} and REM telescopes both initiated automatic slews to the burst, resulting in optical observations in the R and UV bands (1700-6000 \AA, with the \textit{Swift} UltraViolet-Optical Telescope, UVOT) beginning at $T_0+51$ s and $T_0+68$ s, respectively. The \textit{Swift} X-ray Telescope (XRT) began observing the burst at $T_0+51$ s, providing time-resolved spectroscopy in the 0.3-10 keV band.   Subsequent optical spectroscopy by Gemini-N and the Hobby-Eberly Telescope (HET) confirmed the redshift of $z=0.937$ \citep{2008Natur.455..183R}.\\
Finally, this burst was fortuitously located within the field of view of the Milagro observatory. Although no evidence for emission was found in the Milagro data, upper limits on the flux above 10 GeV were derived \citep{2012ApJ...753L..31A}.  
\subsection{GRB 130427A}
On 2013 April 27  one of the most energetic bursts,  GRB 130427A,  was observed from radio wavelengths to  GeV gamma rays.  GRB 130427A triggered the Gamma-ray Burst Monitor (GBM)  onboard the Fermi satellite at $T_0$=07:47:06.42 UTC \citep{2013GCN..14473...1V}. The Large Area Telescope (LAT) followed-up this burst until it became eclipsed by the Earth 715 s  after the GBM trigger.  In addition to a bright peak at $T_0\sim$ 15~s,  this burst displayed the highest fluence with isotropic energy of $\sim 1.4\times 10^{54}$ erg and the highest energy photons ever detected, 73 GeV and 95 GeV  observed at 19 s and 244 s, respectively \citep{2014Sci...343...42A}.   Rapid Telescope for Optical Response \citep[RAPTOR;][]{2010SPIE.7737E..23W}      reported on a bright optical flash which was temporally correlated with the LAT peak. This optical flash had  a magnitude of $7.03\pm 0.03$ and was detected in the time interval of [14  - 16 s] after the GBM trigger \citep{2014Sci...343...38V}.\\
BAT triggered on the ongoing burst at 07:47:57.51 UTC, and  UVOT and XRT started observations at $\sim T_0+181$ s and $\sim T_0+195$s, respectively \citep{2014Sci...343...48M}.  The LC exhibited by the BAT instrument showed a complex structure with a duration of $\sim$ 20 s.  Due to its extremely bright prompt emission, this burst was also detected by other satellites (SPI-ACS/INTEGRAL\citep{2013GCN..14484...1P}   AGILE \citep{2013GCN..14515...1V}, KW \citep{2013GCN..14487...1G}, NuSTAR \citep{2013ApJ...779L...1K}   RHESSI \citep{2013GCN..14590...1S}) and multiple ground- and space follow-up facilities (MAXI/GSC \citep{2013GCN..14462...1K}, VLT/X-shooter \citep{2013GCN..14491...1F}).  Optical spectroscopy from Gemini-North found the redshift of the GRB to be z=0.34 (confirmed later by VLT/X-shooter \citealp{2013GCN..14491...1F}), revealing the closeness to Earth \citep{2013GCN..14686...1L} and the  optical/near infrared (NIR) counterpart observed with the Hubble Space Telescope suggested the association of GRB 130427A with a Type Ic supernova (SN2013cq;  \citealp{2013GCN..14686...1L, 2013ApJ...776...98X}). \\
The high-altitude water Cherenkov observatory (HAWC; \citealp{2013GCN..14549...1L, 2015ApJ...800...78A}) followed up this burst and although  GeV - TeV photons were not detected, upper limits in the flux were derived.\\
\subsection{Constraints provided by TeV Observatories}

\subsubsection{Milagro Observatory}
Before, during and after the prompt phase of GRB 080319B,  Milagro observatory simultaneously collected  data from the two data acquisition (DAQ) systems, the main and scaler system \citep{2012ApJ...753L..31A}.   The main DAQ system reads out coincident signals and reconstructs the direction and energy of the atmospheric shower events. The scaler DAQ counts the hits in each photomultiplier tube (PMT) and searches for a statistical excess over the background. Data from both DAQ systems were analyzed to get the upper limits on the GeV - TeV $\gamma$-ray flux. For instance, the standard analysis consisted in searching for an excess of events above the background in temporal and spatial coincidence with the main $\gamma$-ray pulse reported by Konus on board the Wind satellite \citep{2008GCN..7482....1G}.   This analysis showed no significant excess of events (30 events collected with a predicted background of 29.7) associated during the main $\gamma$-ray pulse reported by Konus.
\subsubsection{HAWC Observatory}
HAWC with  an order of magnitude better sensitivity and angular resolution  than its predecessor, the Milagro observatory,  could follow-up GRB 130427A  \citep{2015ApJ...800...78A}.  This burst took place under disadvantageous conditions for HAWC observation (it was running 10\% of the final detector).   Based on the trigger time, HAWC selected  eight different time periods to search for photons in the energy range of 0.5 GeV - 1 TeV.  In the selected periods,  no statistically significant excesses  were found and upper limits were placed. These upper limits were converted to integral flux upper limits using the HAWC effective area for the declination of GRB130427A.
\subsection{Comparison: GRB 080319B and GRB 130427A}
Given some similarities,   we summarise in  Table~1  the relevant observational quantities for GRB 080319B and GRB 130427A.  
\begin{center}
\begin{center}
\scriptsize{\textbf{Table 1. Observed quantities for  GRB 080319B and GRB 130427A.}}\\
\end{center}
\begin{tabular}{ l c c }
  \hline \hline
 \scriptsize{Parameter} & \scriptsize{GRB080319B} & \scriptsize{GRB130427A} \\
 \hline
\scriptsize{Isotropic energy ($\times 10^{54}$ erg)}    & \scriptsize{1.3}  &  \scriptsize {1.2} \\
\scriptsize{Redshift}  & \scriptsize{0.937}  &  \scriptsize{$0.34$ }\\
\scriptsize{Period of peak correlations (s)$^{(b)}$ }    & \scriptsize{$2.75 - 57.0$}  &  \scriptsize{$9.31 - 19.31$}\\
\scriptsize{$L^{(a)}_{\rm x}/L_{\rm op}$}    & \scriptsize{$15.3$ }  &  \scriptsize{ $89.1$}\\
\scriptsize{Low-energy power-law index ($\alpha$)}    & \scriptsize{ $0.833$ }  &  \scriptsize{$ 0.789$}\\
\scriptsize{High-energy power-law index ($\beta$)}    & \scriptsize{ $3.499$ }  &  \scriptsize{$ 3.06$}\\
\scriptsize{Spectral break energy ($E^{\rm obs}_{\rm p}$)}    & \scriptsize{ $ 651$ }  &  \scriptsize{$ 830$}\\
\hline
\end{tabular}
\end{center}
\begin{flushleft}
\scriptsize{
{(a) The X-ray luminosities were reported by Swift-BAT \citep[Band 15 - 350 keV;][]{2014Sci...343...48M}. The magnitude of the optical flashes was $5.3$ in the V-band and $7.03$ in the R-band for GRB080319B  \citep{2008Natur.455..183R} and GRB130427A \citep{2014Sci...343...38V}, respectively.\\ 
  (b) They are taking from \citet{2008Natur.455..183R} and \cite{2014Sci...343...38V} for GRB080319B and GRB130427A, respectively.  These corresponds to the periods where optical flashes and the $\gamma$-ray components shows close correlations. }
 }
\end{flushleft}
\section{Emission Processes}
To study the optical flash we show a model based on the IS and the early-afterglow ES.   The convention $Q_x=Q/10^x$ is used  in c.g.s. units with  the universal constants   c=$\hbar$=1 in natural units.
\subsection{Internal shocks}
In the standard fireball model, inhomogeneities in the jet lead to internal shell collisions occurring at ${\small r_j=2\Gamma^2\,t_\nu}$,  with $t_v$ the variability time scale of the engine and $\Gamma$ the bulk Lorentz factor.  The total energy density $U=1/(8\,\pi\,m_p)\,\Gamma^{-4}\,L_j\,t^{-2}_{\nu}$ in the internal shock is equipartitioned to amplify the magnetic field and to accelerate particles through the microphysical parameters $\epsilon_B$ and $\epsilon_e$, respectively.  $L_j$ is the isotropic equivalent kinetic luminosity and m$_p$ is the proton mass.  Once the magnetic field is amplified,  relativistic electrons are efficiently cooled down  via synchrotron radiation and IC scattering.\\
Taking into account the bright optical and the MeV $\gamma$-ray components, it is naturally thought that synchrotron and IC radiation could describe the optical flash and the MeV $\gamma$-ray component, respectively \citep{2000ApJ...544L..17P, 2008MNRAS.384...33K}.  Considering this assumption and requiring the  Band function \citep{1993ApJ...413..281B},  then the Compton $Y_1$ parameter, defined as the radio of IC scattering to synchrotron energy losses, is
 {\small
 \be\label{Y1}
 Y_1\simeq\frac{E^{\rm ssc1}_{\rm \gamma,p}\,F_\nu^{\rm ssc1}}{E^{\rm syn}_{\rm \gamma,p}\,F_\nu^{\rm syn}}\,,
 \ee
  }
 where $E^{\rm syn}_{\rm \gamma,p}$ is the synchrotron energy defined by
\be\label{Esyn}
E^{\rm syn}_{\gamma,p}=\frac{3\,q_e}{2\pi\, m_e}\,(1+z)^{-1}\,\Gamma\, B'\,\gamma_e^2\,,
\ee
 with the comoving magnetic field  $B'= \epsilon_B^{1/2}\,\Gamma^{-2}\,L_j^{1/2}\,t^{-1}_\nu$, the peak synchrotron flux 
 \be\label{flux}
F^{\rm syn}_\nu=\frac{m_e\sigma_T}{12\pi q_e}\,(1+z)\,\Gamma\, B'\, N_e\,\,D^{-2}\,,
\ee
and  {\small $E^{\rm ssc1}_{\rm \gamma,p}\simeq \gamma_e^2\,E^{\rm syn}_{\rm \gamma,p}$}. Here,  $m_e$ is the electron mass, $q_e$ is the elementary charge,  $\sigma_T$ is the Thomson cross section, $D$ is the luminosity distance, $N_e$ is the number of radiating electrons and $\gamma_e=\left(\frac{Y_1}{\tau}\right)^{\frac12}$ is the electron Lorentz factor \citep{2009ApJ...692L..92Z} with $\tau$ the optical thickness of the source to Thomson scattering given  by
\be\label{tau}
\tau=\frac{\sigma_T\,N_e}{4\pi R^2}\,.
\ee
 From eqs. (\ref{Y1}), (\ref{Esyn}), (\ref{flux}) and (\ref{tau}), the emitting radius is
 \be
R=\sqrt{\frac{9q^2_e}{2\pi m_e^2}}(1+z)^{-1}\,D\,\gamma_e^2\, (F_\nu^{\rm ssc1} E^{\rm ssc1}_{\rm \gamma,p})^{1/2}\,Y^{-1}_1\,[E^{\rm syn}_{\rm \gamma,p}]^{-1}\,.
\ee
Due to the MeV $\gamma$-ray emission is stronger than the optical flash, then a third spectral emission arising from the second-order IC scattering would be expected \citep{2008MNRAS.391L..19K}.   
The second inverse Compton scattering takes place just above the Klein-Nishina (KN) limit, where the electron scattering cross-section is $\sim 0.4\,\sigma_T$.  The KN suppression becomes important only at {\small $E_{KN}> 132.9\left(\frac{10\,{\rm eV}}{E^{\rm syn}_{\gamma, p}} \right)^{-1/2}\,\Gamma_3\, {\rm GeV}$}.   Hence, the twice-scattered photon takes all the electron energy, then the third spectral component peaks at energies around 
{\small
\be
E^{\rm ssc2}_{\rm \gamma,p}\simeq \frac{[E^{\rm ssc1}_{\rm \gamma, p}]^2}{E^{\rm syn}_{\rm \gamma,p}}\,,
\ee
}
and theoretically, the Compton $Y_2$ parameter is given by \citep{2008MNRAS.391L..19K}
\be\label{Y2_Theo}
Y_2=0.4m_e (1+z)^{-1}\, \Gamma\,E^{\rm ssc1}_{\rm \gamma,p}\,\gamma_e^{-1}\,Y_1\,.
\ee
Observationally,  the Compton parameter of second-order IC scattering can be obtained through
{\small
 \be\label{Y2}
 Y_2\simeq\frac{E^{\rm ssc2}_{\rm \gamma,p}\, F_\nu^{\rm ssc2} }{E^{\rm ssc1}_{\rm \gamma,p}\, F_\nu^{\rm ssc1}}\,,
 \ee
}
where $E^{\rm ssc2}_{\rm \gamma,p}$ is the corresponding Band function peak energy in the GeV - TeV energy range given by  {\small $E^{\rm ssc2}_{\rm \gamma,p}\simeq \gamma_e^2\,E^{\rm ssc1}_{\rm \gamma,p}$} and  $E^{\rm ssc2}_{\rm \gamma,p}\, F_\nu^{\rm ssc2}$ are obtained from the GeV -TeV limits derived by Milagro and HAWC observatories. 
\subsection{Are the HAWC and Milagro upper limits restrictive?}
\begin{figure}[!htbp]
\begin{center}
\includegraphics[width=\columnwidth,angle=0]{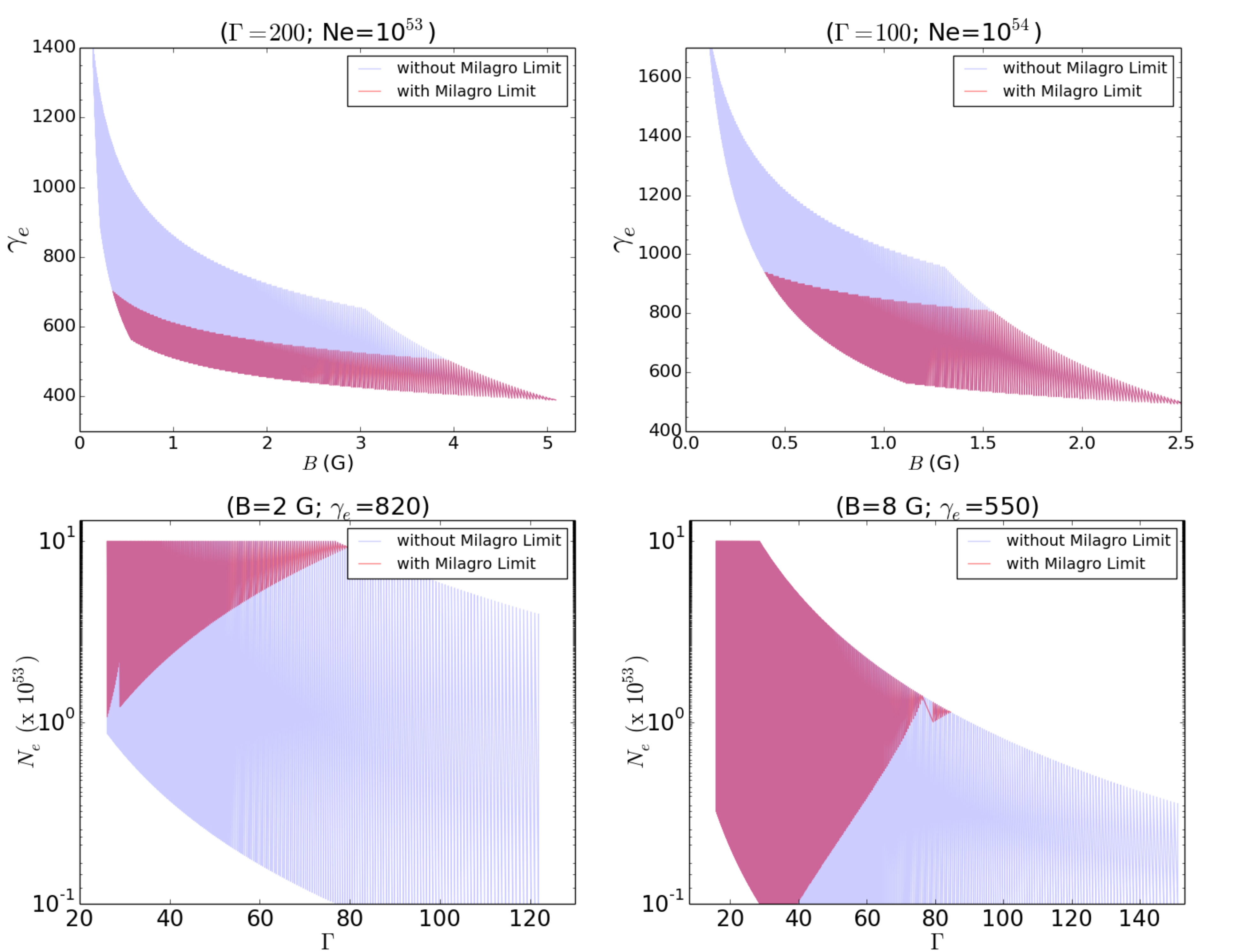}
\caption{Range of parameters found using the  observables of GRB080319B with and without Milagro limits. Upper panels show the electron Lorentz factor as a function of magnetic field and lower panels show the electron density as a function of bulk Lorentz factor. }
\label{par_080319B}
\end{center}
\end{figure}

\begin{figure}[!htbp]
\begin{center}
\includegraphics[width=\columnwidth,angle=0]{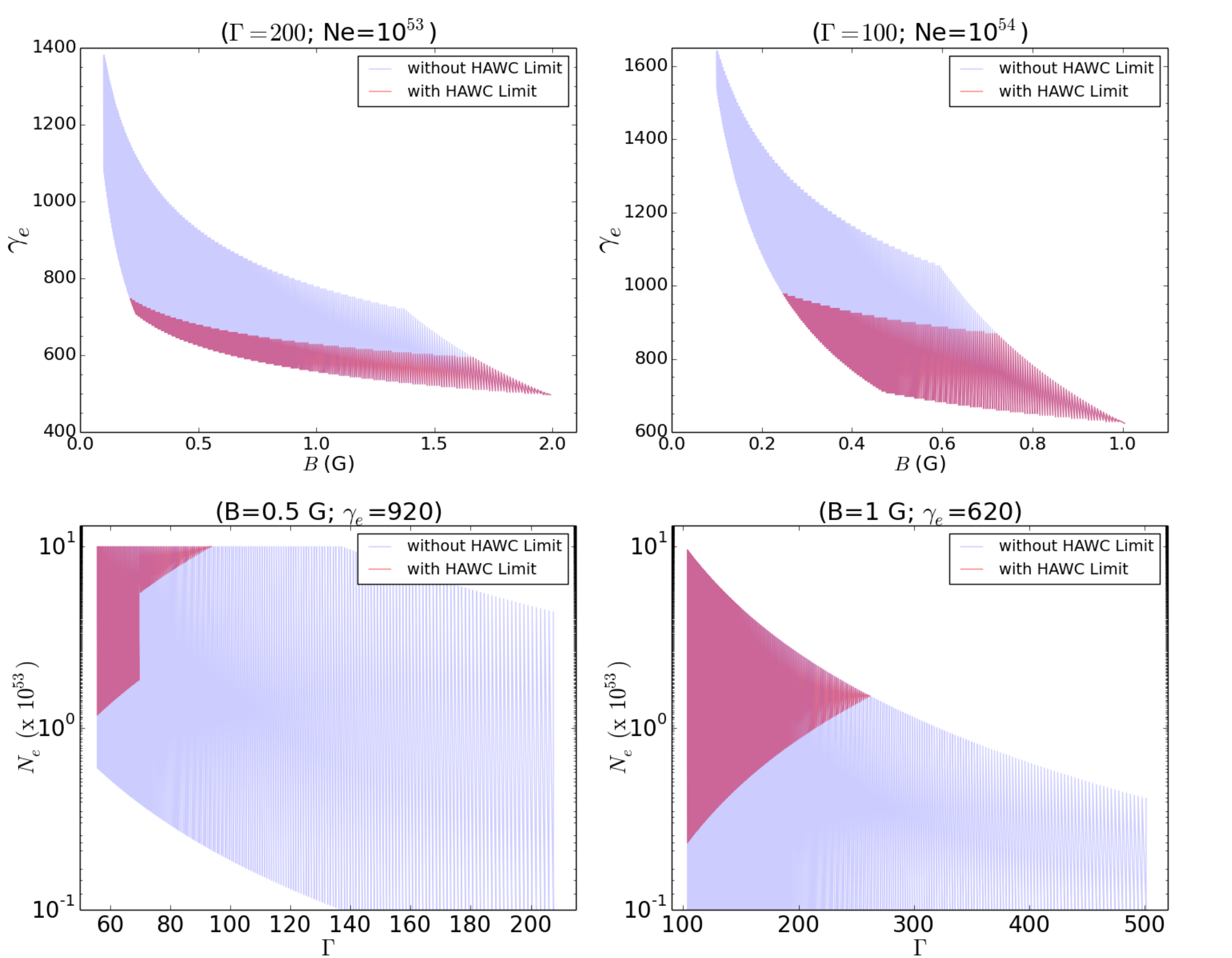}
\caption{Range of parameters found using the  observables of GRB130427A with and without HAWC limits. Upper panels show the electron Lorentz factor as a function of magnetic field and lower panels show the electron density as a function of bulk Lorentz factor. }
\label{par_130427A}
\end{center}
\end{figure}

The upper limits set by Milagro and HAWC $\gamma$-ray observatories in the range of tens of GeV to 1 TeV are used  for GRB080319B and GRB130427A, respectively, in order  to constrain the range of parameters of electron Lorentz factor ($\gamma_e$), bulk Lorentz factor ($\Gamma$),  number of radiating electrons ($N_e$) and magnetic field ($B$).  The restrictions of these parameters are given using  the second-order IC scattering originated in internal shocks  which was described above.\\
Figures  \ref{par_080319B}  and \ref{par_130427A} show the range of parameters found using the observables of GRB080319B and GRB 130427A with (red color) and without (blue color) Milagro and HAWC upper limits, respectively, with the following assumptions: i) the optical flash peaks at {\small $0.5 \leq E_{\rm opt}\leq 1.5\,{\rm eV}$} with a corresponding flux in the range of {\small $ 10^{-8}\leq \,F_{\rm \nu,opt}\leq10^{-7}\, {\rm erg\, cm^{-2}\, s^{-1}}$}, and ii) the $\gamma$-ray emission peaks at {\small $0.1 \leq E_{\rm \gamma}\leq 0.6\,{\rm MeV}$} with a corresponding flux in the range of  {\small $5\times10^{-6}\leq \,F_{\nu,\gamma}\leq 5\times10^{-5}\, {\rm erg\, cm^{-2}\, s^{-1}}$}.\\ 
\\
\\
Upper panels in Figure \ref{par_080319B} show the electron Lorentz factor as a function of magnetic field.  Left-hand panel exhibits that electron Lorentz factor and magnetic field are in the ranges of $400\lesssim \gamma_e\lesssim  1400$ and $0.5\lesssim B\lesssim 5\,{\rm G}$, respectively, for $\Gamma=200$ and $N_e=10^{53}$ and the right-hand panel displays that are in the ranges of $500\lesssim  \gamma_e\lesssim  1600$ and $0.3\lesssim B\lesssim 2.5\,{\rm G}$ for $\Gamma=100$ and $N_e=10^{54}$. Panels show that the electron Lorentz factors larger than $\gamma_e \gtrsim 700$ (left) and  $\gamma_e \gtrsim 950$ (right) are restricted when the Milagro upper limits are taken into consideration.\\
\\

Lower panels in Figure \ref{par_130427A} show the number of radiating electrons as a function of bulk Lorentz factor.  Left-hand panel exhibits that the number of radiating electrons and the bulk Lorentz factor are in the ranges of $0.1\lesssim N_e\lesssim10\times 10^{53}\,$ and  $60\lesssim \Gamma\lesssim 210$ for  $B=0.5\,{\rm G}$ and $\gamma_e=920$, and the right-hand panel displays that are in the ranges of $0.1\lesssim N_e\lesssim 8\times 10^{53}\,$ and  $105\lesssim \Gamma\lesssim 500$  for $B=1\,{\rm G}$ and $\gamma_e=620$.  Panels show that  bulk Lorentz factors larger than $\Gamma\gtrsim 100$ (left) and $\Gamma\gtrsim 250$ (right) are forbidden when the HAWC upper limits are considered.  \\ 
Figures  \ref{par_080319B}  and \ref{par_130427A} show that the HAWC and Milagro upper limits are restrictive when the second-order IC scattering is considered. 
\subsection{Early-afterglow external shocks}
The afterglow emission begins at a distance where most of the energy carried by the outflow is transferred to the circumburst medium \citep{1992MNRAS.258P..41R},  generating forward  and reverse shocks \citep{1994ApJ...430L..93R, 1994MNRAS.269L..41M}.  We will use the subscripts f and r to refer throughout this paper the forward  and reverse shocks, respectively.   In this subsection we are going to adopt the wind afterglow model  $\rho=A r^{-2}$ with {\small $ A=  A_{\star}  (5.0\times 10^{11})\,{\rm g/cm}$} proposed to describe the early multiwavelength emission in GRB130427A  \citep{2016ApJ...818..190F}.\\
\subsubsection{Light curves from Forward shock Emission}
Once the outflow has been accelerated relativistically and has gone into the stratified wind, it begins to be decelerated, leading to a continuous softening of the synchrotron forward-shock spectrum.  The synchrotron spectrum is usually obtained using the deceleration, cooling and acceleration time scales, and the maximum flux given by the peak spectral power \citep[see, e.g.][]{2000ApJ...536..195C}.   Given the synchrotron spectral breaks \citep{2016ApJ...818..190F}, the light curves in the fast and slow-cooling regime are obtained through the synchrotron spectrum.  The light curves in the fast-cooling regime is  
\begin{eqnarray}
\label{fcsyn_t}
[F_{\rm \nu,f}]^{\rm syn}= \cases{ 
F_{\rm \nu,fl}\,t_1^{-\frac{1}{4}}    \,\,\,,\hspace{1cm} E^{\rm syn}_{\rm c,f}<E_\gamma<E^{\rm syn}_{\rm m,f}, \cr
F_{\rm \nu,fh}\,t_1^{-\frac{3p-2}{4}}\,\,\,,\hspace{0.5cm}E^{\rm syn}_{\rm m,f}<E_\gamma<E^{\rm syn}_{\rm max,f}\,, \cr
}
\end{eqnarray}
where  $F_{\rm \nu,fh}$ is
\bary\label{A1}
F_{\rm \nu,fh}&=&  2.8\times 10^{-1}\,\,  {\rm mJy}\,k_f^{-1}(1+z)^{\frac{p+2}{4}}\xi^{3(1-\frac{p}{2})} \,  \epsilon_{e,f}^{p-1}\cr
&&\hspace{1.5cm}\times\,\epsilon_{B,f}^{\frac{p-2}{4}}\,E_{54.7}^{\frac{p+2}{4}}\,D_{28}^{-2} \,\left(\frac{E_{\rm \gamma}}{100\,{\rm MeV}}\right)^{-\frac{p}{2}}\,, 
\eary
\noindent and $F_{\rm \nu,fl}$ is given in \cite{2015ApJ...804..105F}. The terms $k_f=(1+Y_f)$ and $\xi$  are parameters defined in \cite{2000ApJ...536..195C} and  \cite{2016ApJ...818..190F} and $E$ is the equivalent kinetic energy.  The light curve in the slow-cooling regime is
\begin{eqnarray}
\label{scsyn_t}
[F_{\rm \nu,f}]^{\rm syn}=\cases{
F_{\rm \nu,sl}\,t_1^{-\frac{3p-1}{4}}\,\,\,,\hspace{1cm} E^{\rm syn}_{\rm m,f}<E_\gamma<E^{\rm syn}_{\rm c,f},\cr
F_{\rm \nu,sh}\,t_1^{-\frac{3p-2}{4}}\,\,\,,\hspace{1cm} E^{\rm syn}_{\rm c,f}<E_\gamma<E^{\rm syn}_{\rm max,f}\,, \cr
}
\end{eqnarray}
with $F_{\rm \nu,sh}$ and $F_{\rm \nu,sl}$ given by
\bary\label{A2}
F_{\rm \nu,sh}&=&  1.9\times 10^5\,\, {\rm mJy}\,k_f^{-1}(1+z)^{\frac{p+2}{4}}\,\xi^{3(1-\frac{p}{2})}\epsilon_{e,f}^{p-1}\,  \epsilon_{B,f}^{\frac{p-2}{4}} \cr
&&\hspace{2.5cm}\times\,E_{54.7}^{\frac{p+2}{4}}\,D_{28}^{-2} \,\left(\frac{E_{\rm \gamma}}{10\,{\rm keV}}\right)^{-\frac{p}{2}}\,,  
\eary
and
\bary\label{A3}
F_{\rm \nu,sl}&\simeq& 7.4\times 10^7\,\, {\rm mJy}\,(1+z)^{\frac{p+5}{4}}\xi^{\frac{(1-3p)}{2}} \epsilon_{e,f}^{p-1}\,\epsilon_{B,f}^{\frac{p+1}{4}}\,A_{\star}\cr
&&\hspace{2.5cm}\times\, E_{54.7}^{\frac{p+1}{4}}\,D_{28}^{-2}\, \left(\frac{E_{\rm \gamma}}{2\,{\rm eV}}\right)^{\frac{1-p}{2}}\,,
\eary
respectively. The transition time between fast- to slow-cooling regime occurs at {\small $t^{\rm syn}_0=2.3\times10^7 {\rm s}\,\left(\frac{1+z}{1.34}\right)\xi_{-0.3}^{-4}\epsilon_{e,f}\,\epsilon_{B,f}\,A_{\star}$}.
\subsubsection{Light curves from Reverse shock emission}
Synchrotron light curves are derived in \cite{2000ApJ...545..807K}.  The synchrotron fast-cooling regimen is  \citep{2003ApJ...597..455K}
\begin{eqnarray}
\label{rcsyn_t}
[F_{\rm \nu,r}]^{\rm syn}\propto \cases{ 
t^{\frac12}\,\,\,,\hspace{1.1cm} t<t_d, \cr
 t^{-3}\,\,\,,\hspace{1cm} t>t_d\,, \cr
}
\end{eqnarray}
where $t_d$ is the crossing time. The synchrotron flux at the deceleration time is give by   
\bary\label{synpeak}
F^{\rm syn}_{\rm \gamma,peak,r}&\simeq&  2.3\times 10^4  {\rm  mJy}\,  (1+z)^{5/4}\,k_r^{-1}\,\xi^{\frac12}   \epsilon_{B,r}^{-\frac14}\,\Gamma^{-1}_{2,r}\,A_{\star}^{-\frac12} \,\cr
&&\hspace{1.9cm}\times\,D_{28}^{-2}\, E^{\frac54}_{54.7}\,t_{d,1}^{-\frac34}\,  \left(\frac{E_{\rm \gamma}}{2\,{\rm eV}}\right)^{-1/2}\,.
\eary
The light curve of Compton scattering emission is analytically derived in \cite{2015ApJ...804..105F}. It is written as 
\begin{eqnarray}
\label{rcsyn_t}
[F_{\rm \nu,r}]^{\rm ssc}\propto \cases{ 
t^{\frac12}\,\,\,,\hspace{1.2cm} t<t_d, \cr
 t^{-\frac{p-1}{2}}\,\,\,,\hspace{0.7cm} t>t_d\,. \cr
}
\end{eqnarray}
It is worth noting that the decay index of the emission for  $t  > t_d$ might be higher than $\frac{p-1}{2}$ due to the angular time delay effect \citep{2003ApJ...597..455K}.  The SSC  flux peaks at 
\bary\label{sscpeak}
F^{\rm ssc}_{\rm \gamma,peak,r}&\simeq& 2.1\times 10^{-2}  {\rm  mJy}  (1+z)^{-\frac12}\xi^9\,Y_r\,k_r^{-5}\epsilon_{e,r}\,\epsilon_{B,r}^{-\frac72}\,\Gamma^{-6}_{2,r} \, \cr
&&\hspace{1.2cm}\times\,A^{-6}_{\star}\,D^{-2}_{28}\,E_{54.7}\,t_{d,1}^{-\frac12}\left(\frac{E_{\rm \gamma}}{100\,{\rm MeV}}\right)^{-\frac12}.
\eary
\section{Discussion}
%
%
We have introduced two scenarios for the origin of the early optical flashes present in GRB 080319B and GRB 130427A.  In the IS scenario, the optical flash and the MeV $\gamma$-ray emission are correlated through the SSC model whereas in the early-afterglow ES scenario the optical flash is studied together with the multiwavelength afterglow observations.\\  
%
%
\subsection{Internal shocks}
\begin{figure}[!htbp]
\begin{center}
\includegraphics[width=\columnwidth,angle=0]{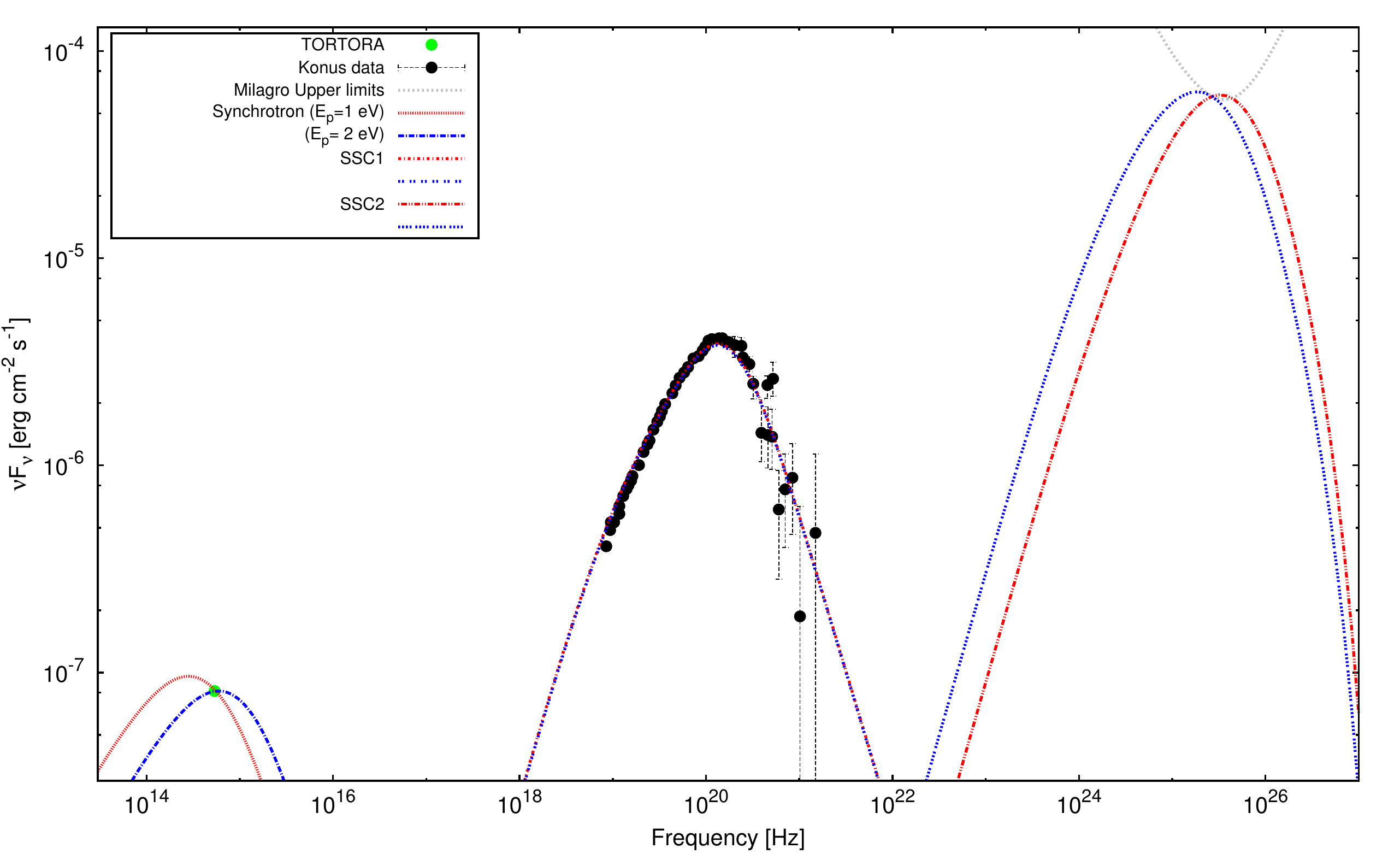}
\includegraphics[width=\columnwidth,angle=0]{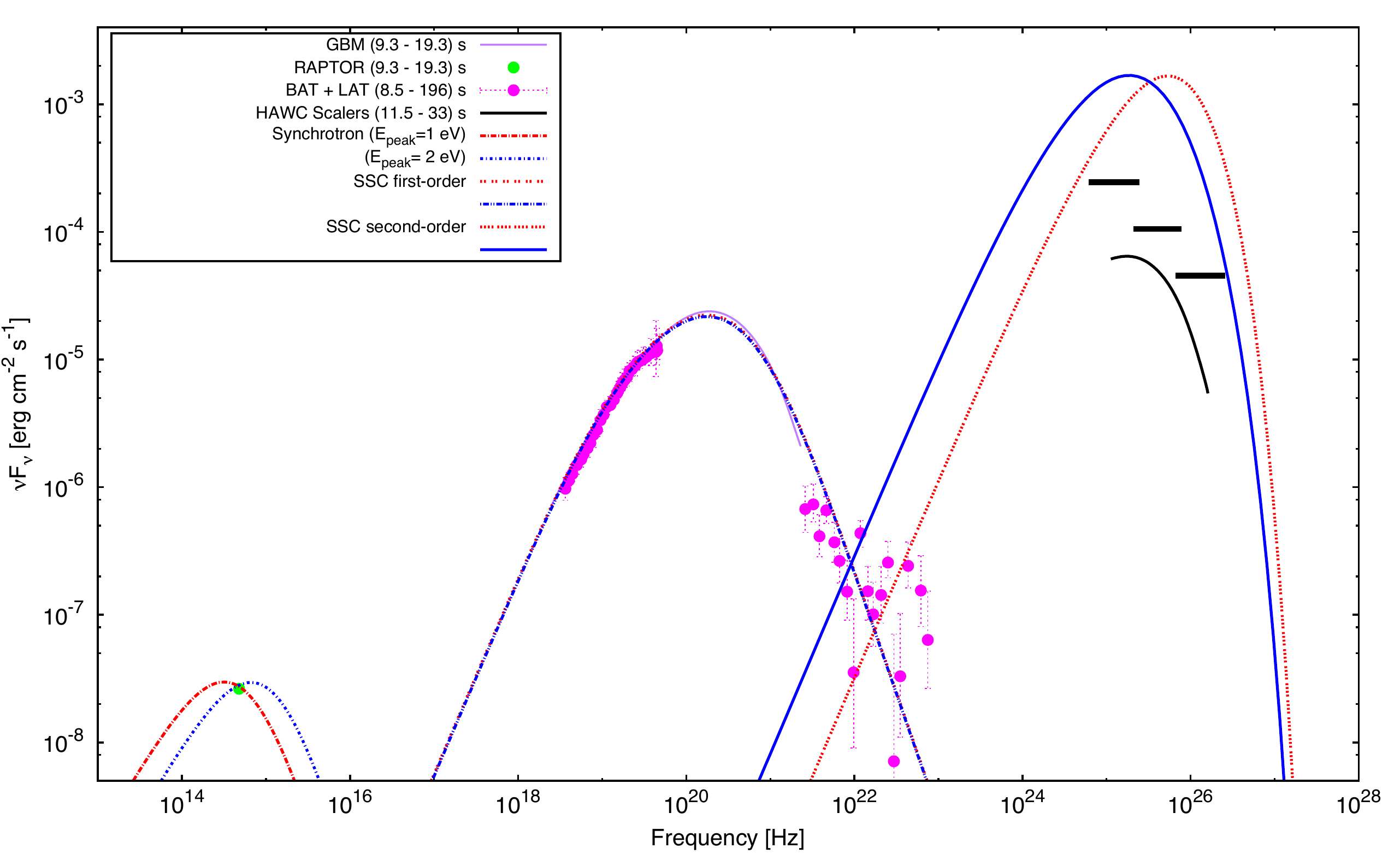}
\caption{Fits of the SED of GRB 080319B (above) and GRB 130427A (below) observations with SSC model of first-  and second-order.  Synchrotron emission has been used to describe the optical data and the IC scattering has been used to fit MeV $\gamma$-ray data. We require the upper limits placed by the Milagro and HAWC experiments, respectively, to constrain the IC scattering of second order. (For details see  \cite{2012ApJ...753L..31A, 2015ApJ...800...78A}).}
\label{fig1}
\end{center}
\end{figure}

 In this scenario, we have assumed that the bright optical flash can be interpreted as synchrotron radiation and the MeV $\gamma$-ray component as the IC scattering of the synchrotron photons by the same population of electrons.  We have required the Band function \citep{1993ApJ...413..281B} to interpret the optical flash and to fit the  $\gamma$-ray component. \\ 
Requiring the parameter values of the Band function ($\alpha$, $\beta$ and $E^{ssc1}_{\gamma,p}$) for GRB 0800319B  \citep{2008Natur.455..183R} and GRB 130427A \citep{2013GCN..14473...1V}   that describe the MeV $\gamma$-ray components,  we compute the values of $\gamma_e$ that describe the optical data for the synchrotron peak in the range  $0.7 < E^{\rm syn}_{\rm \gamma,p}< 1.4\, {\rm eV}$ as shown in Table 2.    Additionally, we obtain the values of the the optical thickness, the strength of  comoving magnetic field, emitting radius and the the number of radiating electrons. Because of MeV $\gamma$-ray component has a large amount of photons,  the radiation process given by the second-order IC scattering must be considered \citep{2008MNRAS.391L..19K}.   From the values obtained (see Table 2) after describing the optical flashes and MeV $\gamma$-ray components with synchrotron and IC scattering emissions, respectively,  we compute that the second-order IC scattering peaks in the energy range where Milagro and HAWC observatories are sensitive. Although these TeV experiments did not collect statistically significant excess of counts, upper limits were placed on the GeV - TeV energies. Considering these upper limits and eqs. (\ref{Y1}) and (\ref{Y2}) we have obtained the Compton parameters for the first and second IC scattering.  Figure 1 shows the fit of the SED of GRB 080319B (above) and GRB 130427A (below) observations with SSC model of first- and second-order.   For the GeV $\gamma$-ray fluxes, we have used the effect of the extragalactic background light absorption modelled in \cite{2008A&A...487..837F}.   As shown in Figure \ref{fig1}, the upper limits set by Milagro and HAWC experiments are useful to constrain the values of the Compton parameters of the second scattering.\\
Table 2 shows the values of $Y_2$ obtained from the theoretical model proposed in  \cite{2008MNRAS.391L..19K} and from the upper limits set by Milagro and HAWC experiments. The theoretical value of $Y_2$ was calculated with $\Gamma=$500 which was estimated from the variability time scale and the inferred burst radius.   Comparing both the theoretical and observational values, one can see that the values of $Y_2$ obtained with the theoretical model are forbidden, thus indicating that SSC scenario is disfavored to explain the correlation of optical and the MeV $\gamma$-ray emission.  This result is consistent with the fact that in the SSC framework, IC scattering must have less fluctuations than the synchrotron photon field, however, the MeV gamma-ray LC shows a higher variability that the optical counterpart.   The resulting in a much lower Compton parameter for the second scattering than the first one (see Table 2) is due to that the KN suppression does not affect the first scattering but affect the second.  From eqs. (\ref{Y2_Theo}) and (\ref{Y2}), and  using the upper limits derived by Milagro and HAWC observatories, we found that the values allowed of the bulk Lorentz factor for $E^{\rm syn}_{\rm \gamma,p}$= 0.7 (1.4) eV are  $\Gamma \lesssim$ 180 (167) for GRB 080319B, and  $\Gamma \lesssim$ 77 (124) for GRB130427A.\\
\begin{center}
\begin{center}
\scriptsize{\textbf{Table 2. Parameters found after fitting the multiwavelength observations with an internal shock model.}}
\end{center}
\begin{tabular}{ l c c c c}
  \hline \hline
 \scriptsize{} & \multicolumn{2}{c}{\small{GRB 080319B}}   &\multicolumn{2}{c}{\small {GRB 130427A}} \\
 \hline
 {\small Synchrotron radiation}\\
\hline
\scriptsize{$E^{\rm syn}_{\rm \gamma,p}\,\, ({\rm eV})$ }          & \scriptsize{$0.7$} & \scriptsize{$1.4$} &  \scriptsize {$0.7$} &  \scriptsize {$1.4$} \\
\scriptsize{$\gamma_e$}          & \scriptsize{$10^{2.98}$} & \scriptsize{$10^{2.83}$}  &  \scriptsize {$10^{3.03}$} &  \scriptsize {$10^{2.88}$} \\
\scriptsize{$Y_1$}    & \scriptsize{ 101.2}  & \scriptsize{$ 100.7$} &  \scriptsize {$297.9$} &  \scriptsize {$299.7$} \\
\scriptsize{$B'\,\,(G)$}          & \scriptsize{$0.5$} & \scriptsize{$2.1$} &  \scriptsize {$0.3$} &  \scriptsize {$1.2$} \\
\scriptsize{$\tau\,\,(\times 10^{-4})$ }          & \scriptsize{$1.1$} & \scriptsize{$2.2$} &  \scriptsize {$2.7$} &  \scriptsize {$5.1$} \\
\scriptsize{$ R\,\,(\times 10^{15}\,\,\rm cm)$}    & \scriptsize{$11.8 $} & \scriptsize{$2.9  $} &  \scriptsize {$2.8$} &  \scriptsize {$7.4$} \\
\scriptsize{$N_e\,\,(\times 10^{53})$}          & \scriptsize{$2.9$} & \scriptsize{$0.3$} &  \scriptsize {$39.5 $} &  \scriptsize {$5.3$} \\
 \hline
 {\small SSC1}\\
 \hline
\scriptsize{$\alpha$}    & \scriptsize{$ 0.833$}& \scriptsize{$ 0.833$}  &  \scriptsize {$0.789$} &  \scriptsize {$0.789$} \\
\scriptsize{$\beta$}     &   \scriptsize{$3.499$} & \scriptsize{$ 3.499$}   &  \scriptsize{$3.06$} &  \scriptsize{$3.06$} \\
\scriptsize{$E^{\rm ssc1}_{\rm \gamma,p}\,\,({\rm keV}$) }          & \scriptsize{$ 651$} & \scriptsize{$651$}  &  \scriptsize {$830$} &  \scriptsize {$830$} \\
 \hline
{\small SSC2}\\
 \hline
\scriptsize{$Y_2$ (Theor.)}    & \scriptsize{6.1} & \scriptsize{$12.3$} &  \scriptsize {$25.3$}&  \scriptsize {$35.6$} \\
\scriptsize{$Y_2$ (Observ.)}    & \scriptsize{$\lesssim2.2$} & \scriptsize{$\lesssim4.1$} &  \scriptsize {$\lesssim3.9$}&  \scriptsize {$\lesssim8.8$} \\
\scriptsize{$E^{\rm ssc2}_{\rm \gamma,p}\,\, ({\rm GeV})$ }        & \scriptsize{$ 605.4$}  & \scriptsize{$302.7 $}  &  \scriptsize {$984.1$} &  \scriptsize {$492.1$} \\
 \hline
\end{tabular}
\end{center}
\begin{center}
\end{center}
On the other hand, using the upper limits on the prompt optical emission, \cite{2009MNRAS.393.1107P} showed that under general conservative assumption the inverse Compton scattering mechanism suffers from an ``energy crisis",  which is the overproduction of a very-high-energy component that would carry much more energy than the observed MeV prompt.  Authors explored the parameter space to see whether there exists a regime for less energy in the second-order IC component than in the MeV $\gamma$-ray prompt.  They found that the parameter space for $\Gamma$ and $\gamma_e$ is limited to a very small region.    In our work, upper limits are set at TeV - GeV range instead of optical band previously developed in \cite{2009MNRAS.393.1107P}.  Therefore, using  the fit of the optical and MeV $\gamma$-ray data, we found that the flux ratio is $\frac{F_\nu^{\rm ssc1}}{F_\nu^{\rm syn}}\approx10^{-3}$ which is much smaller than that analysis performed in \citealp{2009MNRAS.393.1107P} ( $\frac{F_\nu^{\rm ssc1}}{F_\nu^{\rm syn}}=10^{-2}$)  and \citealp{2007ApJ...669.1107Y} ($\frac{F_\nu^{\rm ssc1}}{F_\nu^{\rm syn}} \gtrsim 0.1$). Considering the condition {\small $E^{\rm ssc2}_{\rm \gamma,p}\, F_\nu^{\rm ssc2}    \lesssim      E^{\rm ssc1}_{\rm \gamma,p}\, F_\nu^{\rm ssc1}$} ($Y_2\lesssim 1$), we found that the values allowed of the bulk Lorentz factors  for $E^{\rm syn}_{\rm \gamma,p}$= 0.7 (1.4) eV lie at $\Gamma \lesssim$ 8.2 (4.1) for GRB 080319B and $\Gamma \lesssim$ 20 (14)  for GRB130427A.\\
We show that the Lorentz factors calculated using the upper limits set by Milagro and HAWC observatories are much less than that obtained by theoretical considerations  \citep{2008MNRAS.391L..19K}.  This inconsistency illustrates that optical flashes are not correlated with the $\gamma$-ray components, and hence produced likely by different electron populations.\\
%
%
\subsection{Early-afterglow External shocks}
In this scenario, we use the early-afterglow ES model  presented in \cite{2016ApJ...818..190F}  in oder to fit the early optical flashes together with the multiwavelength data observed in GRB 0803189B and GRB 130427A.   From the deceleration time scales $t_{\rm dec}=$10 s (GRB 130427A) and 50 s (GRB 080319B),  the values of bulk Lorentz factors are $\Gamma_{\rm f}=520$ and $\Gamma_{\rm f}=550$ for stellar densities  $A=10^{10}\,{\rm g/cm}$ and $5\times10^{10}\,{\rm g/cm}$, respectively.\\

\begin{figure}[!htbp]
\begin{center}
\includegraphics[width=\columnwidth]{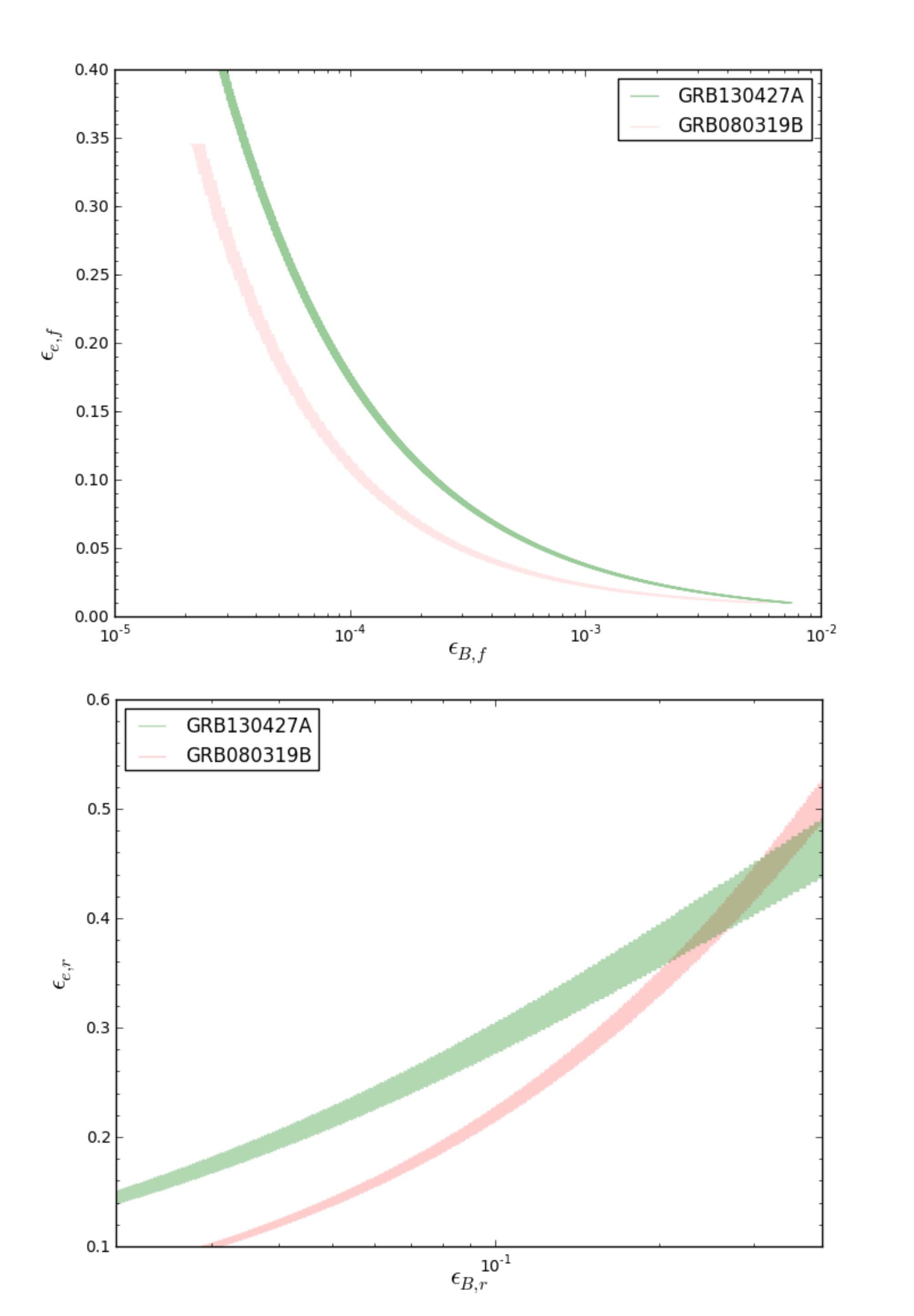}
\caption{The microphysical parameter space that describe the temporally extended emissions (panel above) and the brightest optical flashes (panel above) present in GRB 080319B and GRB 130427A.}
\label{fig3}
\end{center}
\end{figure}

\begin{figure}[!htbp]
\begin{center}
\includegraphics[width=\columnwidth,angle=0]{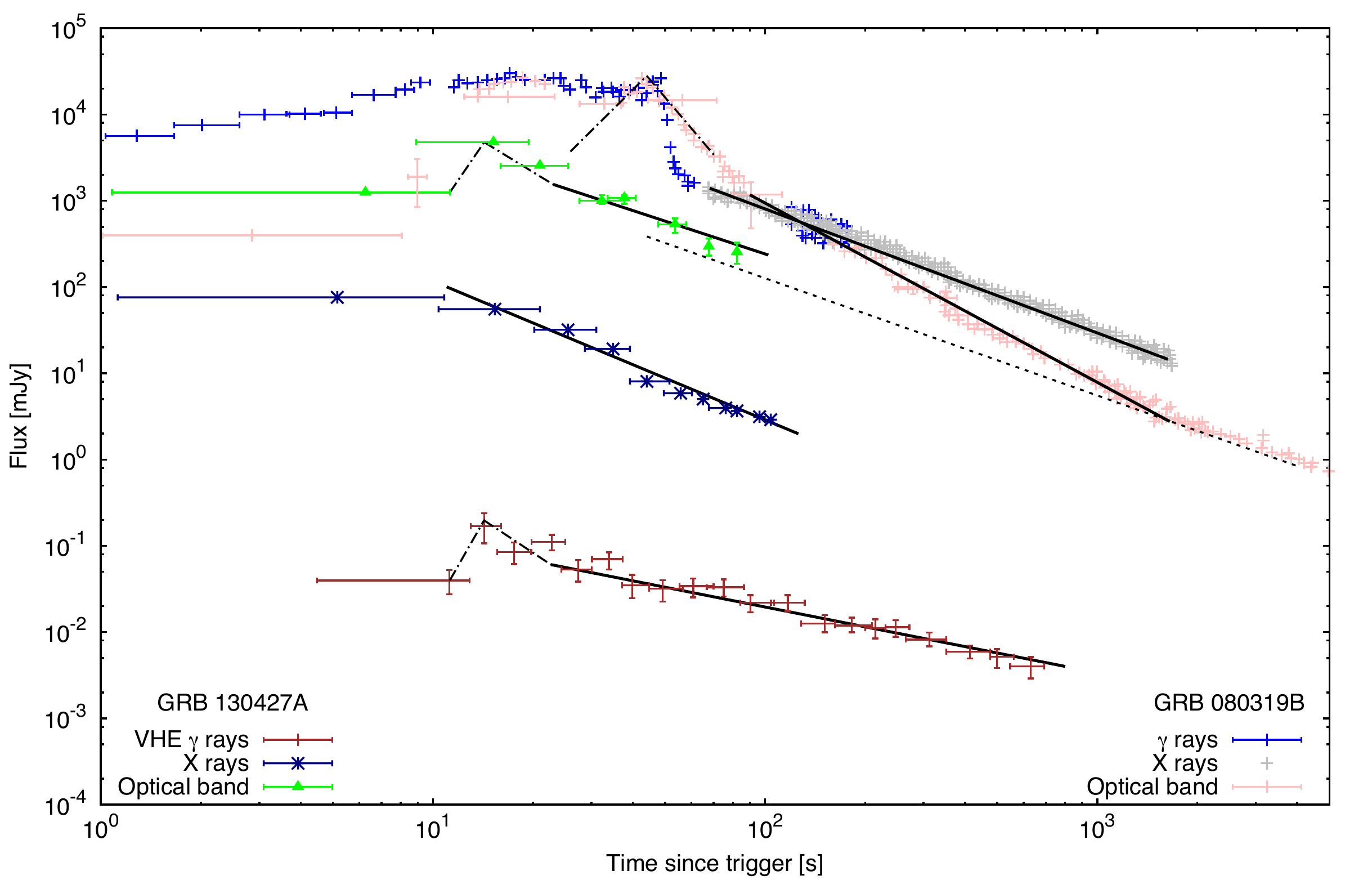}
\caption{$\gamma$-ray, X-ray and optical light curves of GRB 080319B and GRB 130427A. The combine X-ray and Swift-BAT data are extrapolated down into the XRT energy range (0.3 - 10 keV). }
\label{fig4}
\end{center}
\end{figure}

In order to obtain the values of microphysical parameters,  the optical flashes and  the multiwavelength data are fitted using  the Chi-square $ \chi^2$ test \citep{1997NIMPA.389...81B}.  Figure \ref{fig3} (upper panel) displays  the microphysical parameter space ($\epsilon_{B,f}$, $\epsilon_{e,f}$)  that reproduce the long-lived emissions for p=2.2.
The value of the power index of the electron distribution was obtained through the closure relation of synchrotron flux ($F_\nu\propto t^{-\alpha}\nu^{-\beta}$) with  the observed slopes of temporal decays of X-ray ($\alpha_{X}=1.31\pm0.03$; \citep{2008MNRAS.391L..19K}) and optical ($\alpha_{3,Opt}=1.25\pm0.02$; \citealp{2008Natur.455..183R}) fluxes for GRB 080319B and GeV $\gamma$-ray ($\alpha_{GeV}=-1.17\pm0.06$; \citealp{2014Sci...343...42A}), X-ray ($\alpha_{X}=-1.29^{+0.02}_{-0.01}$; \citealp{2014Sci...343...48M}) and optical ($\alpha_{opt}=-1.67\pm 0.07$; \citealp{2014Sci...343...38V}) fluxes for GRB 130427A.  The optical flux of GRB 080319B afterglow at $10^3$ s was extrapolated to earlier times considering that it was eclipsed for the optical flux from RS  \citep{2008MNRAS.391L..19K}.   Lower panel exhibits the microphysical parameter space ($\epsilon_{B,r}$, $\epsilon_{e,r}$), that describe the bright optical flashes. Although we adjust the bright LAT peak in GRB 130427A \citep[see][]{2016ApJ...818..190F}, the  microphysical parameter region found for describing this LAT peak is not included.  It was done to display the similarities between the parameter spaces for both bursts.  The extended fluxes have been fitted using the  synchrotron radiation from FS  and the bright peaks  with  synchrotron  radiation from RS.  The optical fluxes with power index $\alpha_{2,Opt}=2.24\pm0.03$ \citep{2008MNRAS.391L..19K} can be explained with the LC of synchrotron emission from RS.   From the values of the quantities obtained and reported in Table 4 can be seen that synchrotron emission describing the optical flash evolves in the fast cooling regime  {\small $E^{syn}_{\rm \gamma,m,r} > E^{syn}_{\rm \gamma,c,r}$} and  for $t \gtrsim 60\, {\rm s}$, it becomes in the slow cooling regime.  Following  \cite{2013NewAR..57..141G}, the synchrotron flux for $E^{\rm syn}_{\rm \gamma,m,r}< E^{\rm syn}_{\rm \gamma,r}< E^{\rm syn}_{\rm \gamma,c,r}$ behaves as $F_\nu\propto \,t^{-\frac{3(5p+1)}{16}}\,E^{-\frac{p-1}{2}}$, thus reproducing the observed temporal index $\frac{3(5p+1)}{16}=2.5$.\\ 
\begin{center}
\begin{center}
\scriptsize{\textbf{Table 3. Parameters found after fitting the multiwavelength data with the early-afterglow ES model.}}
\end{center}
\begin{tabular}{ l c c }
  \hline \hline
 \scriptsize{} & \scriptsize{GRB 080319B} & \scriptsize{GRB 130427A} \\
 \hline
 Forward shock\\
\hline
\scriptsize{$\epsilon_{B,f}$}    & \scriptsize{$5\times 10^{-5}$}  &  \scriptsize {$3\times 10^{-5}$} \\
\scriptsize{$\epsilon_{e}$}    & \scriptsize{$0.3 $}                         &  \scriptsize{0.32} \\
\scriptsize{$A\, (10^{10}\,{\rm g/cm})$ }    & \scriptsize{$1$}  &  \scriptsize {$5$} \\
\scriptsize{$\Gamma_f$}    & \scriptsize{$520$}  &  \scriptsize {550} \\
 \hline
 Reverse shock\\
 \hline
\scriptsize{$\epsilon_{B,r}$}    & \scriptsize{$0.15$}  &  \scriptsize {$0.13$} \\
\scriptsize{$\epsilon_{e}$}    & \scriptsize{$0.3$}  &  \scriptsize {0.32} \\
\scriptsize{$A\, (10^{10}\,{\rm g/cm})$}    & \scriptsize{$1$}  &  \scriptsize {$5$} \\
\scriptsize{$\Gamma_r$}    & \scriptsize{$700$}  &  \scriptsize {550} \\
 \hline
\end{tabular}
\end{center}
\begin{center}
\end{center}
In Table 3, we summarise the microphysical parameters, the stellar wind densities and the bulk Lorentz factors found after fitting the multiwavelength data from GRB 130427A and GRB 080319B.  Computing the magnetisation parameter, one can see that 
it lies in the range ($0.1\leq\sigma\leq1$) which is consistent with the description of the bright peak from RS and the duration of shock crossing time shorter than $T_{90}$ \citep{2003ApJ...595..950Z,2004A&A...424..477F, 2017arXiv171008514F, 2016ApJ...831...22F}.  Otherwise, when the GRB outflow  crossed the RS, it would have been suppressed \citep{2004A&A...424..477F, 2005ApJ...628..315Z}.\\ 
 Using the values of parameters reported in Table 3, the observable quantities have been computed, as shown in Table 4. In this Table can be observed some features:   i) Comparing the values of synchrotron spectral breaks from RS, one can see that the synchrotron spectrum lies in the fast cooling regime for both  GRB080319B ($E^{\rm syn}_{\rm \gamma,c,r} \lesssim   E^{\rm syn}_{\rm \gamma,m,r}$) and  GRB130427A  ($E^{\rm syn}_{\rm \gamma,c,r} \ll  E^{\rm syn}_{\rm \gamma,m,r}$).   Once the period where the optical and the $\gamma$-ray components exhibited close correlations have finished,  the synchrotron spectrum of GRB 080319B changes from fast to slow cooling regime ($E^{\rm syn}_{\rm \gamma,m,r} \lesssim   E^{\rm syn}_{\rm \gamma,c,r})$ producing a temporal power index of 2.5 and the synchrotron spectrum of GRB 080319B keeps in the fast cooling regime.  It explains the different behaviour of the optical flux after the optical flashes.    ii) The values of the characteristic SSC energies  ($E^{\rm ssc}_{\rm \gamma,m,r}$) illustrate  that whereas a peak at $\sim$ 100 MeV can be detected in GRB 130427A, just a peak at much lower energies $\sim$ 2 MeV can be observed in  GRB 080319B.   iii) From the strength of magnetic fields derived in the forward- and reverse-shock regions can be seen that the ejecta of both bursts are magnetised. \\ 
 Figure \ref{fig4} shows the contributions of synchrotron radiation from FS and RS  to the multiwavelength afterglow observed in GRB 080319B and GRB 130427A.\\
\begin{center}
\begin{center}
\scriptsize{\textbf{Table 4. Observable quantities obtained  with the parameters reported in Table 2 and the ES model.}}\\
\end{center}
\begin{tabular}{ l c c }
  \hline \hline
 \scriptsize{} & \small{GRB080319B} & \small{GRB130427A} \\
 \hline\hline
 Forward shock\\
\hline\hline
\scriptsize{$t_{\rm dec}$ (s)} & \scriptsize{$22.5$} & \scriptsize{$9.9$}  \\
\scriptsize{$B'_f$ (G)} &\scriptsize{$0.5$} &\scriptsize{$18.9$}\\
\hline
\small{Synchrotron emission}\\
\hline
\scriptsize{$E^{\rm syn}_{\rm \gamma,a,f}$ (eV)}    & \scriptsize{$2.7\times 10^{-4}$}  &  \scriptsize{$3.3\times 10^{-2}$} \\
\scriptsize{$E^{\rm syn}_{\rm \gamma,m,f}$ (keV)}    & \scriptsize{$1.3$}  &  \scriptsize{$23.1$} \\
\scriptsize{$E^{\rm syn}_{\rm \gamma,c,f} $} (eV)   & \scriptsize{245.6}  &  \scriptsize{1.3} \\
\scriptsize{$E^{\rm syn}_{\rm \gamma,max,f}$ (GeV)}    & \scriptsize{$6.5 $}  &  \scriptsize{$107.7$} \\
\hline
\small{SSC emission}\\
\hline
\scriptsize{$E^{\rm ssc}_{\rm \gamma,m,f}$ (TeV)}    & \scriptsize{$0.4 $}  &  \scriptsize{$22.1$} \\
\scriptsize{$E^{\rm ssc}_{\rm \gamma,c,f}$ (TeV)}    & \scriptsize{$0.4$}  &  \scriptsize{$1.4\times 10^{-7}$} \\
\scriptsize{$E^{\rm KN}_{\rm \gamma,f}$ (GeV)}    & \scriptsize{$5.4\times10^3$}  &  \scriptsize{$102.3$} \\
 \\
 \hline\hline
 Reverse shock\\
\hline\hline
\scriptsize{$\Gamma_c$} & \scriptsize{321.8} &\scriptsize{236.7} \\
\scriptsize{$B'_r$ (G)} &  \scriptsize{$21.2$}& \scriptsize{$1.7\times 10^3$}\\
\hline
\small{Synchrotron emission}\\
\hline
\scriptsize{$E^{syn}_{\rm \gamma,a,r}$ (eV)}    & \scriptsize{$1.41\times 10^{-9}$}  &  \scriptsize{$0.5\times 10^{-7}$} \\
\scriptsize{$E^{syn}_{\rm \gamma,m,r}$ (eV)}    & \scriptsize{$1.2 $}  &  \scriptsize{$14.3$} \\
\scriptsize{$E^{syn}_{\rm \gamma,c,r}$ (eV)}    & \scriptsize{$0.3$ }  &  \scriptsize{$2.5\times 10^{-5}$} \\
\hline
\small{SSC emission}\\
\hline
\scriptsize{$E^{ssc}_{\rm \gamma,m,r}$ (MeV)}    & \scriptsize{$1.8$}  &  \scriptsize{$61.2$} \\
\scriptsize{$E^{ssc}_{\rm \gamma,c,r}$ (eV)}    & \scriptsize{$11.5\times 10^{3}$}  &  \scriptsize{$1.4\times 10^{-5}$} \\
\scriptsize{$E^{KN}_{\rm \gamma,r}$ (GeV)}    & \scriptsize{$38.3\times10^3$}  &  \scriptsize{$166.2$} \\
\hline
\end{tabular}
\end{center}
\section{conclusions}
%
%
Both GRB 080319B and GRB 130427A are among the most powerful bursts detected in optical and $\gamma$-ray energy band.  The exceedingly bright optical emission peaking with the $\gamma$-ray components at the early phase of these bursts pose challenges in the theoretical models for IS and/or RS.\\ 
%
In the IS framework, the most natural explanation for the spectral energy distribution of GRB 080319B and GRB 130427A is to interpret the optical flashes by synchrotron emission and the MeV $\gamma$-ray photons by IC scattering. Due to the huge amount of MeV $\gamma$-ray photons, the SSC model predicts the existence of a strong peak at hundreds of GeV.  Although no significant excess of counts coming from these bursts  were observed by Milagro and HAWC observatories,  upper limits were used to constrain this model.  We  show that the HAWC and Milagro upper limits are restrictive when the second -order IC scattering model is considered.  From the value of the second-order Compton parameter found using our model and the parameter space allowed for the bulk Lorentz factors, we conclude that the optical and the MeV $\gamma$-ray components are produced by different electron populations, thus disfavouring the IS scenario.  Our analysis was focused on the case that the optical flashes were created within the emitting region that includes the IC scattering ultra-relativistic electrons.  This analysis was  limited to the important implicit assumption that the moving region is homogenous. It is worth noting that very strong inhomogeneities might change this scenery. \\
%
%
In the early afterglow picture, we have used the leptonic model introduced in  \cite{2016ApJ...818..190F} in order to describe the bright optical flashes. These have been interpreted as  the synchrotron FS emission  in the thick-shell case.  Considering that the ejecta propagating into the stellar wind is decelerated early, at $\sim 50$ s and $\sim 10$ s for GRB 080319B and GRB 130427A, respectively,  we found that  the value of the bulk Lorentz factor as required for most powerful  lGRBs  lies in the range ($\Gamma\sim$ 500 - 550) \citep{2012ApJ...755...12V, 2013ApJ...763...71A, 2014Sci...343...42A,2017ApJ...848...15F, 2017ApJ...848...94F}.  The set of parameters has been limited considering the multiwavelength data.   To find the values of microphysical parameters ($\epsilon_{B,f/r}$, $\epsilon_{e}$), we have assumed that these are constant in the description of  the multiwavelength afterglow data.  The long-lived (LAT, X-ray and optical) emissions was modeled with  synchrotron FS radiation and the bright optical flashes with synchrotron RS emission from RS.\\
%
%
%
Since  GRB  080319B and GRB 130427A are the most  energetic bursts observed with  $z\leq\,1.0$, a large amount of target optical photons is created so that hadrons in the outflow can interact efficiently.  Therefore, these bursts represent potential sources to produce neutrinos with energies between TeV - PeV range.  Searches with IceCube telescope for TeV - PeV muon neutrinos were performed around GRB 0800319B   \citep{2009ApJ...701.1721A} and GRB 130427A  \citep{2013GCN..14520...1B} without collecting excess above background.  Some authors have investigated the possible correlation between the lack of neutrinos and the strengths of magnetic fields \citep{2013PhRvL.110l1101Z, 2014ApJ...787..140F}.  If this is true,  the  null result  reported by this neutrino observatory might be interpreted in terms of the levels of magnetisations found in this paper for both burst.\\
%
%
%
Some authors have claimed that the $\gamma$-ray emission detected by LAT during the prompt phase has an internal origin similar to the  optical counterpart \citep{2011MNRAS.415...77M, 2011ApJ...730..141Z, 2011ApJ...733...22H,  2011ApJ...730....1L}. However, by completing the analysis with the upper limits reported by Milagro and HAWC observatories we have shown that the optical flashes and the $\gamma$-ray components are not produced by the same electron population.    Therefore,  it is  overwhelming evidence that the bright optical flashes comes from the RS as has been explained in this work.   It is important to highlight  that although no significant excess of counts have been detected from both bursts by these TeV $\gamma$-ray observatories,  nowadays bursts with identical features can be detected by this HAWC experiment \citep{2012APh....35..641A,2014arXiv1410.1536A, 2015arXiv150804120W, 2015arXiv150807325L}.   It is worth noting that although GRB 990123 exhibited a bright optical flash \citep{1999Natur.398..400A}, no correlation with gamma-rays was reported  \citep{1999ApJ...517L.109S,  1999MNRAS.306L..39M, 2003ApJ...597..455K} and no upper limits were placed by TeV observatories.   Therefore, similar bursts could bring to light information on external medium density, bulk Lorentz factors and energy fractions converted to accelerate electron and/or amplify magnetic fields, thus potentially further constraining possible models.\\
\acknowledgments
We thank Dirk Lennarz, Ignacio Taboada,  Fabio de Colle and Anatoly Spitovsky for useful discussions. for useful discussions. This work was supported by PAPIIT-UNAM IA102917 and Fermi grant NNM11AA01A (PV).
%
%
%

%
\end{document}